\documentclass[twocolumn,preprintnumbers,amsmath,amssymb]{revtex4}
\usepackage{graphicx}
\usepackage[strict]{changepage}
\usepackage{paralist}
\newcommand{\be}{\begin{equation}}
\newcommand{\ee}{\end{equation}}
\newcommand{\beq}{\begin{equation}}
\newcommand{\eeq}{\end{equation}}
\newcommand{\bea}{\begin{eqnarray}}
\newcommand{\eea}{\end{eqnarray}}
\newcommand{\eml}{\end{mathletters}}

\begin{document}

\title{Disentangling the pair and quartet condensates}

\author{V.V. Baran$^{1,2,3}$}
\email[Email address: ]{virgil.baran@theory.nipne.ro}
\author{D.S. Delion$^{1,3,4}$}

\affiliation{$^1$``Horia Hulubei'' National Institute of Physics and 
Nuclear Engineering, \\ 
30 Reactorului, RO-077125, Bucharest-M\u agurele, Rom\^ania \\
$^2$ Department of Physics, University of Bucharest,
405 Atomi\c stilor, POB MG-11, Bucharest-M\u agurele, RO-077125, Rom\^ania \\
$^3$Academy of Romanian Scientists,
54 Splaiul Independen\c tei, RO-050094, Bucharest, Rom\^ania \\
$^4$Bioterra University, 81 G\^arlei, RO-013724, Bucharest, Rom\^ania}

\begin{abstract}

We study the nontrivial interplay of the well known \emph{pairing} and the more complex \emph{quarteting} correlations in the particular case of $N>Z$ atomic nuclei. Within the new Analytical Disentangled  Condensate model, by implementing the notion of \emph{fractional degeneracy} we obtain the clear physical picture of a rather weakly interacting mixture of quartet and neutron pair condensates which mainly feel each other's influence through Pauli blocking. The basic idea of our approach may be generalized  in order  to scrutinize the extent to which similar manifestations are present in various many-body systems.

\end{abstract}

\maketitle

After more than 60 years the pairing correlations, so pervasive in condensed matter, nuclear \cite{50bcs} and particle physics \cite{njl}, are still actively investigated. In nuclear physics in particular, the richness of many body effects is substantially influenced by presence of two fermion species, protons and neutrons. The existence of four spin-isospin possible combinations allows the formation of strongly correlated  four-particle structures, known as \emph{quartets}, due to the nuclear attractive force. As a result, the $\alpha$-particle is characterized by a large binding energy. In heavy nuclear systems, this structure survives as an $\alpha$-cluster,
as can be seen from the binding energies. In condensed matter systems, conceptually related manifestations may be identified, for example the formation and condensation of bi-excitons in semiconductors \cite{Noz82} or the existence of a quartet superfluid phase in a system of fermionic cold atoms trapped in a one-dimensional optical lattice (see Ref. \cite{Cap07} and references therein).
In nuclear systems, $\alpha$-particles can appear only at relative low nuclear 
densities \cite{Rop98} on the nuclear surface of $\alpha$-decaying nuclei \cite{Del13}. From a theoretical viewpoint, the main difficulty is connected to strong antisymmetrization
effects between nucleons entering $\alpha$-like structures. On the one hand, coordinate space approaches like the THSR ansatz \cite{Toh01} have been successfully applied only to light nuclei or to heavy nucleus+$\alpha$ systems, e.g. $^{212}$Po=$^{208}$Pb+$\alpha$ \cite{Xu17}. On the other hand, configuration space approaches based on correlated quartet structures were recently proven to describe very precisely the four body correlations  induced by the residual nuclear interaction \cite{San12a,danielphd,San12,Neg14,San14,San15,Neg17,
Neg18,Sam15,Fu13,Fu15}, in both $N=Z$ and $N>Z$ nuclei. A unified microscopic description of real space $\alpha$ clustering and configuration (shell model) space quartet correlations is an open problem in theoretical nuclear physics.


Moreover, in $N>Z$ nuclei, it is necessary to consider the interplay of quartet and neutron pair correlations. In this work, we aim to provide a clearer physical picture of such systems by developing new theoretical many body methods based on recent advances in the analytical description of pairing and quarteting correlations \cite{qcmanalitic}. We consider $N$ neutrons and $Z$ protons moving outside a self-conjugate inert core which interact through a charge-independent pairing force. The corresponding isovector pairing Hamiltonian is applicable to both spherical and deformed nuclei,
\beq
\label{ham}
H=\sum_{i=1}^{N_\text{lev}}\epsilon_i N_{i,0}+\sum_{\tau=0,\pm1}\sum_{i,j=1}^{N_\text{lev}}V_{ij}P^{\dagger}_{i,\tau } P_{ j,\tau}~,
\eeq
where $i,j$ denote the single particle doubly-degenerate states and $\epsilon_i$ refers to the single particle energies; a time conjugated state will be denoted by $\bar{i}$. The $N_{i,0}$ operator counts the total number of particles,
$N_{i,0}=\sum_{\tau=\pi,\nu}\left(c^\dagger_{i,\tau}c_{i,\tau}+c^\dagger_{\bar{i},\tau}c_{\bar{i},\tau}\right)$, whereas the isovector triplet of pair operators is given by $P^\dagger_{i,1}=c^\dagger_{i,\nu}c^\dagger_{\bar{i},\nu},
P^\dagger_{i,-1}=c^\dagger_{i,\pi}c^\dagger_{\bar{i},\pi},
P^\dagger_{i,0}=\dfrac{1}{\sqrt{2}}\left(c^\dagger_{i,\nu}c^\dagger_{\bar{i},\pi}+c^\dagger_{i,\pi}c^\dagger_{\bar{i},\nu}\right)$. 

We adopt the Quartet Condensation Model framework, where first one defines a set of collective $\pi\pi$, $\nu\nu$ and $\pi\nu$ Cooper pairs $\Gamma^\dagger_\tau(x)\equiv\sum_{i=1}^{N_\text{lev}}x_i P^{\dagger}_{i,\tau}$, 
which depend on a set of mixing amplitudes $x_i$, $i=1,2,...,N_{\text{lev}}$. A collective quartet operator is then constructed by coupling two collective pairs to the total isospin $T=0$, $Q^\dagger(x)\equiv\left[\Gamma^\dagger\Gamma^\dagger\right]^{T=0}_{S=0}\equiv 2\Gamma^\dagger_1(x)\Gamma^\dagger_{-1}(x)-\big(
\Gamma^\dagger_0(x)\big)^2$. 
In the case of $N=Z$ nuclei, the ground state of the Hamiltonian (\ref{ham}) may be described as a ``condensate'' of such $\alpha$-like quartets, $| \Psi_{q}(x) \rangle=\big(Q^{\dagger}(x)\big)^{n_q}|0\rangle$, where $n_q$ is the number of quartets. For $N>Z$ nuclei it turns out that, in addition to the quartet condensate, a pair condenstate accounting for the excess neutrons must be considered \cite{San12}. In this case the ground state may be written as $| \Psi(x,y) \rangle=\left[Q^{\dagger}(x)\right]^{n_q}\left[\Gamma^\dagger_1(y)\right]^{n_p}|0\rangle$,
where the number of excess neutron pairs is $n_p=(N-Z)/2$ and $n_q$ is the maximum number of quartets that may be constructed, $n_q=(N+Z-2n_p)/4$. The concept of a ``condensate'' is used here to denote the state obtained by acting with the same operator a number of times on the vacuum, and should not  be confused with an ideal boson-type condensate.

 By construction, the above state has a well defined particle number and isospin, the latter being defined by the excess neutrons, $T=n_p$. Its structure is defined by the mixing amplitudes $x_i$ and $y_i$, which are determined numerically by the minimization of the Hamiltonian expectation value subject to the unit norm constraint, i.e. 
\be
\nonumber
\delta\langle \Psi (x,y)| H| \Psi(x,y)\rangle=0~,~~ \langle  \Psi(x,y) | \Psi (x,y)\rangle=1~.
\ee
There currently exist two equivalent approaches used in order to compute these quantities. The method proposed in \cite{San12, San12a} makes use of the recurrence relations obeyed by the matrix elements of the pair, number of particle and isospin operators in the auxiliary basis $|n_1 n_2 n_3 n_4\rangle=[\Gamma_{1}^{\dagger}(x)]^{n_1}[\Gamma_{-1}^{\dagger}(x)]^{n_2}[\Gamma_{0}^{\dagger}(x)]^{n_3}[\Gamma_{1}^{\dagger}(y)]^{n_4}|0\rangle$ of states having a well defined number of each kind of pairs. The analytical relations method recently proposed in Ref. \cite{qcmanalitic} involves the symbolic evaluation of norm and Hamiltonian average starting from the basic $SO(5)$ algebra of pair, isospin and number of particle operators. We refer the reader to Ref. \cite{qcmanalitic} for a more detailed comparison of the two methods.

In this work we introduce a unified approach which combines the benefits of the previously described procedures. Namely, using the very recently developed computing capabilities of the Cadabra2 computer algebra system \cite{cdb1}, it has been possible to implement symbolically the derivation algorithm   of the recurrence relations (presented e.g. in Refs. \cite{danielphd,Neg18}). They are then employed to obtain the analytical form of the the quantities of interest for the model. This more refined procedure ensures a two order of magnitude improvement in the running times of the symbolic evaluation code, as opposed the basic $SO(5)$ algebra implementation. As a consequence, the significantly more complex cases of $N>Z$ nuclei and of isoscalar pairing may now be easily tackled from an analytical perspective. For simplicity, in this work we limit ourselves to the description of the isovector pairing correlations in $N>Z$ nuclei. 
   Analogously to the case of $N=Z$ nuclei, the norm and the Hamiltonian average as functions of the mixing amplitudes may be expressed as $\langle  \Psi(x,y) | \Psi(x,y)\rangle = \mathcal{N}(x,y)$, $\mathcal{H}(x,y)\equiv\langle  \Psi(x,y) | H| \Psi(x,y)\rangle = E(x,y) +v(x,y)$, for a given number of quartets and pairs. These are polynomial functions which may be conveniently expressed in terms of the sums of powers of the mixing amplitudes $x$ and $y$, i.e. $\Sigma_{\alpha,\beta}=\sum_{i=1}^{N_{\text{lev}}}~ x_i^\alpha ~y_i^\beta$, $\mathcal{E}_{\alpha,\beta}=\sum_{i=1}^{N_{\text{lev}}} ~ \epsilon_i~ x_i^\alpha~y_i^\beta$, $\mathcal{V}_{\alpha, \beta;\gamma, \delta}=\sum_{i,j=1}^{N_{\text{lev}}}~V_{ij} ~x_i^\alpha~y_i^\beta ~x_j^\gamma~y_j^\delta$, and $\mathcal{U}_{\alpha,\beta}=\sum_{i=1}^{N_{\text{lev}}}  ~V_{ii}~ x_i^\alpha~y_i^\beta$. 

As an example, we present  the formula of the interaction term for the $n_q=n_p=1$  case,

\begin{adjustwidth}{1em}{0cm}$
v^{(n_q=1,n_p=1)}=6\mathcal{V}_{0 , 1 ; 0 , 1} (\Sigma_{2 , 0})^2 + 3\Sigma_{4 , 0} \mathcal{V}_{0 , 1 ; 0 , 1}-24\Sigma_{2 , 0} \mathcal{V}_{0 , 1 ; 2 , 1}-10\mathcal{V}_{0 , 1 ; 4 , 1}-3\mathcal{V}_{2 , 1 ; 2 , 1} + 28\Sigma_{2 , 0} \mathcal{U}_{2 , 2} + 5\mathcal{U}_{4 , 2} + 3\Sigma_{0 , 2} \mathcal{U}_{4 , 0} + 12\Sigma_{0 , 2} \Sigma_{2 , 0} \mathcal{V}_{1 , 0 ; 1 , 0}-12\Sigma_{2 , 2} \mathcal{V}_{1 , 0 ; 1 , 0}-24\Sigma_{2 , 0} \mathcal{V}_{1 , 0 ; 1 , 2}-20\mathcal{V}_{1 , 0 ; 3 , 2} + 12\Sigma_{0 , 2} \mathcal{V}_{1 , 0 ; 3 , 0}-12\mathcal{V}_{1 , 2 ; 3 , 0} + 4\Sigma_{3 , 1} \mathcal{V}_{0 , 1 ; 1 , 0} + 4\Sigma_{1 , 1} \mathcal{V}_{0 , 1 ; 3 , 0} + 12\Sigma_{1 , 1} \mathcal{V}_{1 , 0 ; 2 , 1} + 8\Sigma_{1 , 1} \Sigma_{2 , 0} \mathcal{V}_{0 , 1 ; 1 , 0} + 4\mathcal{V}_{1 , 0 ; 1 , 0} (\Sigma_{1 , 1})^2 + 4\Sigma_{1 , 1} \mathcal{U}_{3 , 1}~.$
\end{adjustwidth}

The presence of additional neutron pairs 
leads to much more complex expressions than in the case of states composed just of quartets. For this reason, we limit ourselves to this relation to illustrate our method; the symbolic code is freely available upon request and may be used to compute any case of interest. 

The QCM has been proven to describe with very good precision the competition between the $\alpha$-like four-body correlations and the conventional pairing correlations in asymmetric nuclei with both isovector and isoscalar proton-neutron pairing \cite{San12,Neg18}. In particular, the analysis of the Schmidth number in Ref. \cite{San12} led to the conclusion that the nucleons experience a stronger degree of entanglement in the complete picture of quartet and pair condensates than in the case of just a product state of neutron pair and proton pair condensates. In the following, we shall analyze more in depth the interplay of quartet and neutron pair structures  by trying to give a more definite answer to the questions: 

a) How much do the quartet structures contribute to the total amount of correlations?

b) How much do the neutron pairs contribute?

c) What about the quartet-pair correlations?
  
  \begin{figure}[ht]

 \center

\includegraphics[width=\columnwidth]{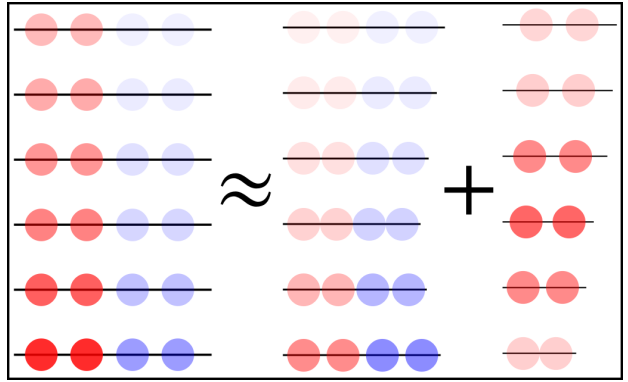}

\caption{ The pictorial representation of our approach: the asymmetric proton-neutron system is disentangled into a condensate of symmetric quartets and a condensate of neutron pairs.  The opacity of each dot is proportional to the average level occupation $\langle n_i \rangle$ for either neutrons (red dots) or protons (blue dots). In the disentangled case, the lenght of each level line is related to the effective fractional degeneracy of the corresponding level, $\Omega_i$ (see also the discussion of Eq. (\ref{effective}) below). They are coupled by the effective Pauli blocking conditions $	\Omega_i^{(\text{quartets})}=1-\langle n_i\rangle^{(\text{pairs})}~,~	\Omega_i^{(\text{pairs})}=1-\langle n_i\rangle^{(\text{quartets})}$.}
\label{fig0}

\end{figure}

\begin{figure*}[ht]

 \centering

\includegraphics[width=\textwidth]{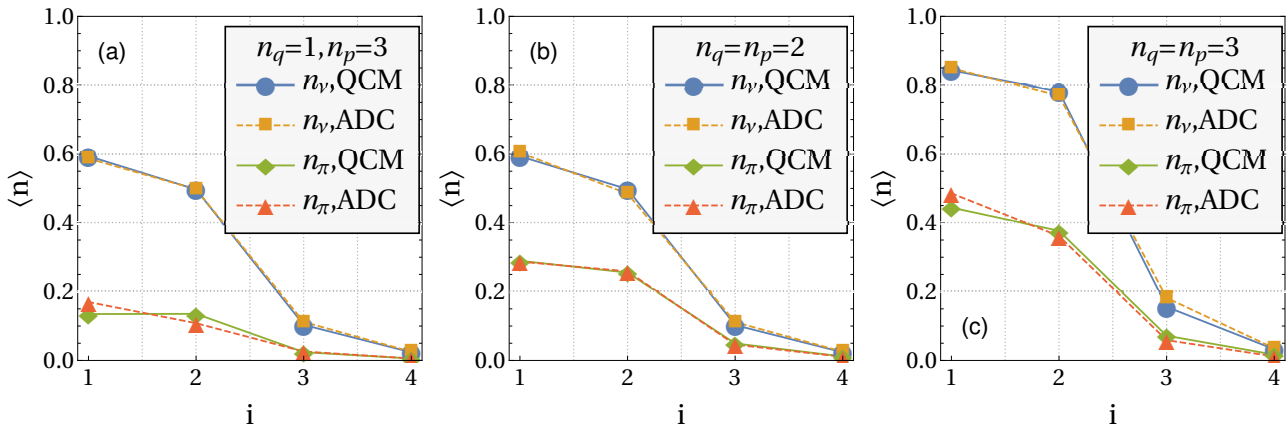}

\caption{Average neutron and proton level occupations $\langle n_{i,\tau}\rangle=\langle N_{i,\tau}\rangle/(2j_i+1)$, $\tau=\pi,\nu$, versus the spherical state index $i$ computed with the Bonn A isovector pairing interaction for the nuclei $^{110}$Te (a),$^{112}$Xe (b) and $^{118}$Ba (c), within the QCM (solid lines) and ADC (dashed lines) approaches.  In the ADC approach, we identify the proton and neutron occupancies as $\langle n_{\pi,i}\rangle\equiv \langle n_i\rangle^{(QCM)}$ and, respectively, $\langle n_{\nu,i}\rangle\equiv \langle n_i\rangle^{(QCM)}+\langle n_i\rangle^{(PBCS)}$.  }
\label{fig1}

\end{figure*}
We shall provide an answer to these questions in the framework of the Analytical Disentangled Condensate model (ADC), described below. Within this approach, a clear separation is performed between the excess neutron subsystem and the subsystem made of quartets. The excess neutron pairs and quartets are considered to be perfectly independent of each other, except for the fact that they live in model spaces whose dimensionality is reduced due to the each other's effective Pauli blocking.
Therefore we treat the excess neutron subsystem within a number projected BCS (PBCS) description and the quartet subsystem within a standard QCM approach for equal numbers of protons and neutrons. We find the ground state of the system in the separable approximation by minimizing the total energy function:
 \beq
 \label{energ_separabil}
 \begin{aligned}
 \mathcal{H}_{ADC}(x,y)=\mathcal{H}^{(QCM)}_{n_q}(x)+\mathcal{H}^{(PBCS)}_{n_p}(y)
 \end{aligned}
 \eeq

 We use the analytical formulas presented in Ref. \cite{qcmanalitic} to compute the quartet contribution $\mathcal{H}^{(QCM)}_{n_q}(x)$ for the QCM case of $n_q$ quartets  and the analogous formulas for the PBCS case (which can be easily derived using the above mentioned symbolic codes) in order to evaluate the excess neutron pair term $\mathcal{H}^{(PBCS)}_{n_p}(y)$, with the same isovector pairing interaction matrix elements in both cases.

 The main point of ADC  is that we consider the two subsystems as interacting only through each other's Pauli blocking. To achieve this, we consider the generalization of the previously defined sums of mixing amplitudes to the case of degenerate levels. If a number of levels $\Omega_i$ share the same energy $\epsilon_i$, and furthermore the interaction matrix elements are equal within each degenerate subspace, the corresponding generalizations for the $\Sigma$, $\mathcal{E}$, $\mathcal{V}$ and $\mathcal{U}$ sums are trivial, e.g. $\Sigma_{\alpha}(x;\Omega)=\sum_{i}~ \Omega_i ~x_i^\alpha$ for the standard quarteting case of Ref. \cite{qcmanalitic}. Here $\Omega_i$ is an integer representing the number of distinct (doubly-degenerate) levels $\epsilon_i$.

 In what follows, we take this interpretation a step further and define the analytical continuation of the model also to the case of arbitrary non-integer degeneracy values, of particular interest being the case of \emph{fractional degeneracies} $\Omega_i<1$. Namely, each model (QCM or PBCS) is solved by minimizing the usual energy function (whose analytical expression is derived in the standard rigourous way) with respect to the mixing amplitudes, e.g. $\delta_x \mathcal{H}^{(QCM)}_{n_q}(x;\Omega)=0$, for any given values of the level degeneracies. This allows us to implement mathematically the Pauli blocking effective space reduction for quartets and pairs by demanding that
 \beq
 \label{effective}
 \begin{aligned}
	\Omega_i^{(QCM)}&=1-\langle n_i\rangle^{(PBCS)}~,\\
	\Omega_i^{(PBCS)}&=1-\langle n_i\rangle^{(QCM)}~,
	\end{aligned}
 \eeq
 where the average level occupations are defined by the analytic formulas of $\langle n_i\rangle^{(PBCS)}\equiv\langle PBCS| N_1/2|PBCS\rangle$ and $\langle n_i\rangle^{(QCM)}\equiv\langle QCM| N_0/4|QCM\rangle$ as functions of the respective mixing amplitudes and degeneracies. The problem is then solved self-consistenly, namely for each set of the mixing amplitudes the corresponding pair and quartet effective degeneracies are computed by iterating Eqs. (\ref{effective}) until convergence is achieved. The  model is schematically represented  in  Fig. \ref{fig0}.

 We apply these ideas to the study of the isovector pairing correlations in the nuclei above $^{100}$Sn. We consider the same model space and interaction as in Refs. \cite{San12,San12a}, namely the spherical spectrum $\epsilon_{2d_{5/2}}=0.0$MeV,  $\epsilon_{1g_{7/2}}=0.2$MeV,  $\epsilon_{2d_{3/2}}=1.5$MeV and  $\epsilon_{3s_{1/2}}=2.8$MeV together with the effective Bonn A isovector pairing potential of Ref. \cite{Jen95}. Let us observe in Fig. \ref{fig1} that in all considered cases the separable ADC treatment reproduces remarkably well the rigourous QCM results regarding both proton and neutron average level occupancies, even for a significant number of valence particle (the largest being 18 for $^{118}$Ba). This is a strong argument that the mismatch in correlation energy between the rigorous QCM and the ADC treatments should be interpreted as the correlation energy between the quartet and excess neutron pair subsystems.

\begin{table}
\caption{Correlation energies in the rigourous approach $E_{corr}^{exact}$ and in the separable case $E_{corr}^{ADC}$,  together with the distribution of the correlation energy among quartets and pairs from Eq. (\ref{energ_separabil}) and the quartet-pair correlation energy $E_{corr}^{q-p}=E_{corr}^{exact}-E_{corr}^{ADC}$, versus the number of quartets and pairs. All energies are expressed in MeV.}
\label{tab1}
\begin{tabular}{c c c c c c c c c }
\hline \hline
Nucleus& $n_q$ & $n_p$ & $E_{corr}^{exact}$  & $E_{corr}^{ADC}$ & $E_{corr}^q$ & $E_{corr}^p$ & $E_{corr}^{q-p}$& $\dfrac{E_{corr}^{q-p}}{n_q\cdot n_p}$ \\	[0.5ex]
\hline
$^{108}$Te &1 & 2& 5.96 & 5.37 & 2.81  & 2.56 & 0.54 & 0.295  \\
 $^{110}$Te &1 & 3& 6.64 & 5.77 & 2.29  & 3.48 & 0.87 & 0.29 \\
 $^{112}$Xe& 2 &  2 & 8.24& 7.10& 4.68& 2.42 & 1.14  & 0.285\\ 
 $^{114}$Xe& 2 &  3 & 8.40& 6.73& 3.67& 3.06 & 1.67 & 0.28\\ 
 $^{118}$Ba & 3& 3  & 9.10 & 6.72& 4.16& 2.56 & 2.28 & 0.26 \\
 
\hline
\end{tabular}
\end{table} 
 
  We have summarized in Table I the correlation energies in the rigourous QCM approach $E_{corr}^{exact}$ and in the separable case $E_{corr}^{ADC}$,  together with the distribution of the correlation energy among quartets and pairs from Eq. (\ref{energ_separabil}) and the quartet-pair correlation energy $E_{corr}^{q-p}=E_{corr}^{exact}-E_{corr}^{ADC}$, versus the number of quartets and pairs.  As a general trend, we observe that quartets and excess neutron pairs bring similar contributions in terms of correlation energy per particle. It it however remarkable that the  quartet-pair correlations grow slower than $n_q\cdot n_p$ with increasing mass number. Specifically, our results indicate that for a small particle number, e.g. $^{108}Te$, the system may be approximated with 90\% accuracy by independent quartets and neutron pairs, while for the heavier $^{118}Ba$ this picture is true to about 75\%, as seen from the ratio $E_{corr}^{q-p}/E_{corr}^{exact}$.

In conclusion, we have succeeded in developing a new treatment of many body correlations which provides a very clear physical picture for the $N>Z$ nuclear systems, in the form of a separable approach based on the analytical method for pairing and quarteting correlations. Our results strongly indicate that in the ground state of $N>Z$ nuclei the isovector pairing interaction induces a rather weakly interacting mixture of quartet and neutron pair condensates, which mainly feel each other's influence through Pauli blocking. 

This new kind of disentangling approach has the potential of
greatly simplifying a wide range of physical many-body problems.

It would be extremely interesting to conduct a similar investigation in the framework of the exact solutions for the isovector and standard pairing cases, in terms of the Richardson equations analytically continued to fractional degeneracies, for various pairing regimes. Our approach may also be easily generalized to 
the description of excited states and to the case of combined isovector-isoscalar pairing. The separable treatment allows, by construction, a description of quartet correlations in a boson formalism along the lines of \cite{phb}, which will be the subject of future works.


We are especially grateful to Kasper Peeters, Durham University, for responding promptly to our request of upgrading Cadabra2 and thus making possible the improved analytical method presented in this paper.
This work was supported by the grants of the Romanian Ministry of Research and Innovation, CNCS - UEFISCDI, PN-
III-P4-ID-PCE-2016-0092, PN-III-P4-ID-PCE-2016-0792, within PNCDI III, and PN-19060101/2019.



\begin{thebibliography}{99}
\bibitem{50bcs}{\emph{Fifty years of nuclear BCS, Pairing in Finite Systems}, R. A. Broglia, V. Zelevinsky editors, World Scientific Publishing (2013)}
\bibitem{njl}{ Y. Nambu, G. Jona-Lasinio,  Phys. Review 122 (1) 345-358.,  Phys. Rev. 124(1): 246-254 (1961)}
\bibitem{Noz82} P. Nozieres, D. Saint-James,  J. Physique {\bf 43}, 1133  (1982).
\bibitem{Cap07} S. Capponi, G. Roux, P. Lecheminant, P. Azaria, E. Boulat, S. R. White, Phys. Rev. A 77, 013624 (2008)
\bibitem{Toh01} A. Tohsaki, H. Horiuchi, P. Schuck, G. R\" opke, Phys. Rev. Lett. {\bf 87}, 192501 (2001).
\bibitem{Xu17} C. Xu, G. Ropke, P. Schuck, Z. Ren, Y. Funaki, H. Horiuchi, A. Tohsaki, T. Yamada, B. Zhou, Phys. Rev. C
{\bf 95}, 061306(R) (2017).
\bibitem{Rop98} G. R\"opke, A. Schnell, P. Schuck, and P. Nozieres,
Phys. Rev. Lett. {\bf 80}, 3177 (1998).
\bibitem{Del13} D.S. Delion and R.J. Liotta,
Phys. Rev. C {\bf 87}, 041302(R) (2013).
\bibitem{San12a} N. Sandulescu, D. Negrea, J. Dukelsky, C. W. Johnson, Phys. Rev. C \textbf{85}, 061303(R) (2012).
\bibitem{danielphd} D. Negrea, {\it Proton-neutron correlations in atomic nuclei}, Ph.D. thesis, 2013, https://tel.archives-ouvertes.fr/ tel-00870588/document.
\bibitem{San12} N. Sandulescu, D. Negrea, C. W. Johnson,
Phys. Rev. C {\bf 86}, 041302(R) (2012).
\bibitem{Neg14} D. Negrea, N. Sandulescu,  Phys. Rev. C \textbf{90}, 024322 (2014).
\bibitem{San14} N Sandulescu et al, J. Phys.: Conf. Ser. \textbf{533}, 012018 (2014).
\bibitem{San15} N. Sandulescu, D. Negrea, D. Gambacurta, Phys. Lett. B, {\bf 751}, 348 (2015).
\bibitem{Neg17} D. Negrea, N. Sandulescu, D. Gambacurta, Prog. Theor. Exp. Phys.  073D05 (2017).
\bibitem{Neg18} D. Negrea, P. Buganu, D. Gambacurta,  N. Sandulescu, Phys. Rev. C \textbf{98}, 064319 (2018).
\bibitem{Fu13} G. J. Fu, Y. Lei, Y. M. Zhao, S. Pittel, A. Arima, Phys. Rev. C {\bf 87}, 044310 (2013).
\bibitem{Fu15} G. J. Fu, Y. M. Zhao, and A. Arima, Phys. Rev. C {\bf 91}, 054318  (2015).
\bibitem{Sam15} M. Sambataro, N. Sandulescu, Phys. Rev. Lett. {\bf 115}, 112501 (2015); Phys. Rev. C {\bf 91}, 064318 (2015); Phys. Lett. B {\bf 763}, 151 (2016); Eur. Phys. J. A {\bf 53}, 47 (2017).
\bibitem{qcmanalitic} V.V. Baran, D.S. Delion, Phys. Rev. C {\bf 99}, 031303(R) (2019).
\bibitem{cdb1} 	K. Peeters, hep-th/0701238; Journal of Open Source Software, 3(32), 1118 (2018); \url{https://cadabra.science}
\bibitem{Jen95}M. Hjorth-Jensen, T.T.S. Kuo , Eivind Osnes,
Phys. Rep. {\bf 261}, 125 (1995).
\bibitem{phb} V.V. Baran, D.S. Delion, arXiv:1902.00065 (2019).
\end{thebibliography}
\end{document}